\documentclass{article}
\usepackage[utf8]{inputenc}
\usepackage{spconf}
\usepackage{float}
\usepackage{graphicx}
\usepackage{multirow}
\usepackage{graphicx}
\usepackage{hyperref}
\usepackage{xcolor}
\DeclareRobustCommand{\IEEEauthorrefmark}[1]{\smash{\textsuperscript{\footnotesize #1}}}
\usepackage{enumitem}

\title{End-to-End Lyrics Recognition with self-supervised learning}
\name{Xiangyu Zhang\IEEEauthorrefmark{1}, Shuyue Stella Li\IEEEauthorrefmark{1}, Zhanhong He \IEEEauthorrefmark{3}, Roberto Togneri\IEEEauthorrefmark{3}, Leibny Paola Garcia\IEEEauthorrefmark{1,}\IEEEauthorrefmark{2}}
\address{
\IEEEauthorrefmark{1}Center for Language and Speech Processing, Johns Hopkins University \\
\IEEEauthorrefmark{2}Human Language Technology Center of Excellence, Johns Hopkins University \\
\IEEEauthorrefmark{3}Department of Computer Science, University of Western Australia
}
\begin{document}

\maketitle

\begin{abstract}
Lyrics recognition is an important task in music processing. Despite traditional algorithms such as the hybrid HMM-TDNN model achieving good performance, studies on applying end-to-end models and self-supervised learning (SSL) are limited. In this paper, we first establish an end-to-end baseline for \emph{lyrics recognition} and then explore the performance of SSL models on \emph{lyrics recognition} task. We evaluate a variety of upstream SSL models with different training methods (masked reconstruction, masked prediction, autoregressive reconstruction, and contrastive learning). Our end-to-end self-supervised models, evaluated on the DAMP music dataset, outperform the previous state-of-the-art (SOTA) system by 5.23\% for the dev set and 2.4\% for the test set even without a language model trained by a large corpus. Moreover, we investigate the effect of background music on the performance of self-supervised learning models and conclude that the SSL models cannot extract features efficiently in the presence of background music. Finally, we study the out-of-domain generalization ability of the SSL features considering that those models were not trained on music datasets. 
\end{abstract}


\section{Introduction}
Lyrics recognition is an important task in music processing. In previous studies, several traditional algorithms such as Hidden Markov Models (HMM) and Time Delay Neural Network (TDNN) achieved good performance~\cite{tsai2018transcribing}. A limited number of studies that use traditional end-to-end models for lyrics recognition tasks have not yet achieved strong results~\cite{gupta2020automatic}\cite{basak2021end}. Over the past few years, new developments in speech recognition using end-to-end models have achieved very low error rates in most Automatic Speech Recognition (ASR) benchmarks~\cite{li2021recent}\cite{gulati2020conformer}. Given such a well-performing end-to-end system, we believe that the same approach can be used for lyrics recognition since lyrics recognition is very similar to ASR~\cite{gupta2020automatic}\cite{dabike2019automatic}.

\begin{figure}[h!]
    \centering
    \includegraphics[width=0.48\textwidth]{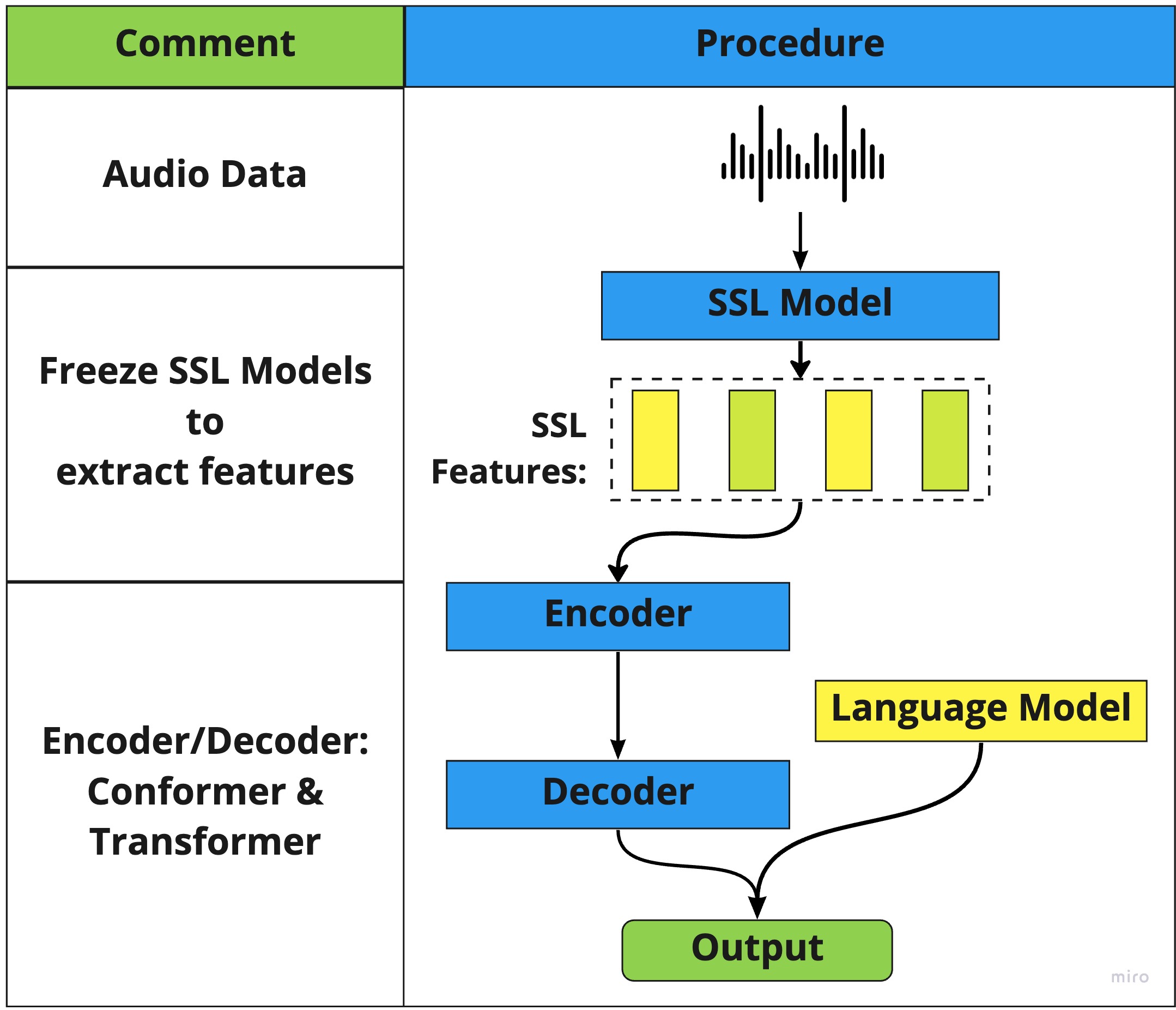}
    \caption{Experiment Pipeline}\label{Pipeline}\vspace{-2mm}
\end{figure}

Automatic sung speech recognition systems typically use the same acoustic features as spoken speech systems. This is motivated by the fact that spoken and sung speech have the same sound production system and that semantic information is conveyed in the same way in both speech styles~\cite{dabike2021use}. However, there are several different acoustic features between sung and spoken speech including pitch ranges, syllable duration, and the existence of vibrato in singing~\cite{dabike2021use}. These differences make sung speech difficult to recognize. Although handcrafted features such as MFCC or LPCs have been shown to achieve a certain degree of effectiveness, these handcrafted features also have performance limitations due to their data-driven nature~\cite{humphrey2018introduction}. For the above reasons, we want to incorporate self-supervised learning models into lyrics recognition task for feature extraction.

SSL models have achieved great success in natural language processing~\cite{MatthewEPeters2018DeepCW}, computer vision~\cite{IshanMisra2020SelfSupervisedLO}, and speech processing~\cite{baevski2020wav2vec,hsu2021HuBERT}. In the field of speech processing, SSL models are often used to effectively extract features and apply them to downstream tasks, such as ASR, speech enhancement and separation, speaker identification, emotion recognition, and other speech tasks~\cite{huang2022investigating}. This paper extends the above approaches into out-of-domain music datasets by integrating SSL models as upstream models (freezing the SSL models and using them to extract features) with end-to-end downstream models in lyrics recognition. Moreover, we study the generalizability of the SSL features, considering that those models were not trained on music datasets.
In summary, this paper has three contributions:
\begin{itemize}[noitemsep,topsep=0pt,leftmargin=10pt]
\item We propose a new lyrics recognition pipeline that extends upstream SSL feature extraction to downstream end-to-end ASR model for music signals with a diverse pitch, tone, and vibrato range.
\item We establish an new baseline for \emph{lyrics recognition} using the DAMP~\cite{DAMP} dataset. The best performance improved by 5.23\% for the dev set and 2.4\% for the test set compared to the previous baseline system even without language model trained by a large corpus. 
\item We analyze the effect of background music on the quality of SSL features to conclude that background music increases the difficulty of feature extraction.
\end{itemize}

\section {Methodology}
In order to investigate the performance of the features extracted by the self-supervised learning model on the lyrics recognition task, we freeze the parameters of the self-supervised learning model, and the specific process is shown in Figure \ref{Pipeline}. Moreover, in order to better compare the performance of the features extracted by the self-supervised learning model on the lyrics recognition task, We also tested the performance of the end-to-end model on the lyrics recognition task without adding the self-supervised learning model.

\subsection{Upstream Self-supervised Feature Extraction}
In this paper, we evaluate four different types of SSL upstream models for lyrics recognition. We selected SSL models based on their training methods. These upstream SSL models can be categorized into \textbf{masked reconstruction model} (Tera~\cite{liu2021tera} and NPC~\cite{liu2020non}), \textbf{masked prediction model} (HuBERT~\cite{hsu2021HuBERT}), \textbf{auto-regressive reconstruction model} (VQ-APC~\cite{chung2020vector}) and \textbf{contrastive model} (wav2vec2.0~\cite{baevski2020wav2vec}). We freeze these SSL models and use them to extract speech representations. Following part of SUPERB's setup~\cite{yang2021superb}, we used the weighted-sum representation from all layers as the final representation F instead of using features from the last hidden layers.
\begin{equation}
    F = \sum_{i=0}^{K-1}w_i F_i,
\end{equation}
\noindent where $K$ is the total number of layers, $F_i$ is the representation extracted from the $i$-th layer, $w_i$ is the weight for the $i$-th layer. The weight vector $\vec{w}$ can be updated during training.

\subsection{Downstream End-to-end Lyrics Recognition}
We use two model architectures to build the downstream end-to-end lyrics recognition model - conformer~\cite{gulati2020conformer} and transformer~\cite{tsunoo2019transformer}. Different upstream SSL feature extraction models work well with different downstream lyrics recognition models. For example, the HuBERT+Conformer architecture achieves a 22\% lower WER than the HuBERT+Transformer architecture.
In order to evaluate the optimal performance of each SSL model, we explore different model combinations and report the best performing combinations.
As a baseline, we use Mel Spectrum as the feature input to the downstream end-to-end models (Transformer and Conformer) to directly observe the effect of SSL feature extractor.

\section{EXPERIMENTAL SETUP}
\subsection{Dataset}
We measure the performance of lyrics recognition on the DAMP dataset~\cite{DAMP}. Previous works focus on the DAMP~\cite{DAMP} and the DALI~\cite{DALI} dataset because of the large amount of data. However, the audio of the DALI dataset needs to be downloaded from streams like YouTube, which results in inconsistent data due to regional restrictions. To achieve reproducible results, this paper establishes an end-to-end lyric recognition benchmark using the DAMP - Sing! 300x30x2 dataset\footnote{we follow the leading Kaldi-based multi-step approach for preprocessing}~\cite{dabike2019automatic}. DAMP - Sing! 300x30x2 dataset contains 149.1 hours of singing clips with no background music. We used the default split(train/dev/test) for the DAMP dataset. There are 66 songs for the validation set and 70 songs for test set. The DAMP dataset is a non-professional recording; most of the singing is mixed with the ambient noise and slightly mismatched with lyrics.

We use the MusDB dataset dataset~\cite{MUSDB18} to explore the impact of background music on the performance of self-supervised learning model. It is a professional corpus made by commercial music and has both versions with and without background music, is the most suitable dataset to explore the impact of background music on the performance of self-supervised learning models. The MusDB dataset consists of 150 full-length audios with background music, including 86 music for the train set, 14 for the validation, and 50 for the test set. MusDB is only 10 hours long due to its limited commercial setting, and the lyrics are manually aligned using phoneme level lyrics alignment \cite{schulze2021phoneme}.

\subsection{Model Architecture and Fine-tuning}
\subsubsection{End-to-End Baseline Model Architecture}\label{sec:baseline}
We test two type of end-to-end models in this paper: transformer~\cite{vaswani2017attention} and conformer~\cite{gulati2020conformer}. For the baseline setup, the best performing models consists of a transformer encoder and transformer decoder. The encoder has 12 blocks and 4 attention heads, 2048 linear units and 0.1 dropout rate; decoder has 6 blocks, 2048 linear units and 0.1 dropout rate. We used the Adam optimizer a learning rate of 1.0 with 25000 warm-up steps. We trained for 100 epochs and evaluated the 10 models with the best performance on the validation set.

\subsubsection{Downstream ASR Model Architecture}\vspace{-2mm}
We use a transformer-based and conformer-based lyrics recognition model that takes the SSL features as inputs. The transformer-based model setup is the same as the baseline described in Section \ref{sec:baseline}. In the conformer-based model, we use a conformer encoder that has 12 blocks and 8 attention heads, 2048 linear units and 0.1 dropout rate, and a transformer decoder that has 8 blocks, 2048 linear units and 0.1 dropout rate.
For the lyrics recognition task , the best performing downstream model for Tera, VQ-NPC and NPC is the transformer-based model. For HuBERT and HuBERT Large, the best performing downstream model is the conformer-based model. 
For training, only 50 epochs were needed with an early termination mechanism. Using the SSL features, the validation loss converges faster than the baseline models with Mel Spectrum features. We use the Adam optimizer with a learning rate of 0.0025 and 40000 warm-up steps.

\subsubsection{Language Model Data and Architecture}\vspace{-2mm}
We examine both LSTM and Transformer language models to generate the final lyrics output. In the LSTM language model, we use 2 layer with a hidden size of 650. In the transformer language model, we use 4 head, 8 layers and a hidden size of 1024. All the data used to train our language model comes from the training set itself, so that no additional information is introduced. When decoding, we use 0.3 as the weight of the language model and use the transformer language model during the final decoding. In the language model step, it is important to note that previous SOTA lyrics recognition systems use a large lyrics language model trained on the LyricsWikia dataset, which is no longer open source \cite{dabike2019automatic}. 

\section{Results and Discussion}
\subsection{Main experiment}
\begin{table}
\centering
\begin{tabular}{| l || c | c |}
 \hline
 SSL Models - WER & Dev set & Test set\\
 \hline\hline
 End-to-End Baseline (Transformer) & 27.0 & 27.6 \\
 Tera + Transformer & 25.1 & 25.2 \\
 NPC + Transformer& 26.0 & 26.7 \\
 HuBERT + Conformer& 21.5 & 21.4 \\
 HuBERT Large + Conformer& \textbf{18.1} & \textbf{17.2} \\
 VQ-APC + Transformer & 25.0 & 26.3 \\
 wav2vec2 + Conformer& 19.7 & 22.4 \\
 wav2vec2 Large + Conformer& 21.2 & 26.2 \\
 \hline
\end{tabular}\vspace{-2mm}
\caption{WER Results on Dsing 30 with Transformer LM}
\label{table:1}\vspace{-2mm}
\end{table}

\begin{table}
\centering
\begin{tabular}{| l || c | c |}
 \hline
 Previous SOTA - WER & Dev set & Test set\\
 \hline\hline
GMM + 4-gram \cite{dabike2019automatic}& 52.95 & 49.50 \\
TDNN-F + 3-gram \cite{dabike2019automatic}& 26.24 & 22.32 \\
TDNN-F + 4-gram\cite{dabike2019automatic}& 23.33 & 19.60 \\
Kaldi LN + VQ + 3-gram \cite{dabike2021use}& not reported &  22.97\\
Kaldi LN + VQ + 4-gram \cite{dabike2021use}& not reported & 19.60 \\
\hline
\end{tabular}\vspace{-2mm}
\caption{SOTA Lyrics Recognition Results on Dsing 30}
\label{table:2}\vspace{-3mm}
\end{table}

In this paper, our experiments are conducted using the ESPNet toolkit~\cite{watanabe2018espnet}. The WER results for Damp dataset is shown in Table \ref{table:1}, while Table \ref{table:2} shows the best results from previous work on the same dataset~\cite{dabike2019automatic}\cite{dabike2021use}. Compared with the strong end-to-end baseline model, even if the parameters of the SSL model are frozen, the performance of all models are improved. 
Previous work on speech enhancement and separation proposed two possible reasons for the disadvantage of SSL models - pretrain-finetune domain mismatch and local information loss due to the focus on long-term dependencies~\cite{huang2022investigating}. However, our results show that the SSL model can significantly improve the performance of the model even with out-of-domain downstream data mismatch (such as vibrato in singing, and multiple singers singing at the same time). This further suggests that the poor performance of SSL models on other non-speech ASR tasks is due to the focus of SSL training to optimize global objectives. 

As listed in Table \ref{table:1}, different SSL models have to be combined with different downstream models to achieve their respective optimal performance. We can see that the optimal downstream model for HuBERT and wav2vec2 is the conformer-based model, and the optimal downstream model of other self-supervised learning models is the transformer-based model. 
In previous studies, research has shown that SSL models with similar training objective, more so than model architecture, tend to produce similar feature representations~\cite{chung2021similarity}, which in turn combines well with similar downstream models.
HuBERT and wav2vec2 were both trained by reconstructing the representation from waveform, which supports to our empirical results that HuBERT and wav2vec2 have the same optimal downstream model. Comparing the HuBERT results with HuBERT Large, and wav2vec2 with wav2vec2 Large, we conclude that the number of parameters in the SSL model is not directly related to the generalization ability of the model features. 
This indicates that during feature extraction, too many parameters could make the SSL model over-fit to the training data, resulting in poor performance on out-of-domain data. Comparing our results in Table \ref{table:1} and SOTA results in Table \ref{table:2}, our models outperform the previous results by 5.23\% on dev set and 2.4\% on the test without the large language model trained on LyricsWikia~\cite{dabike2019automatic,dabike2021use}. We make our results available for future explorations here\footnote{\url{https://github.com/Tonyyouyou/Music-Project-Result}}.

\subsection{Ablation studies}
In this section, we use the HuBERT Large model as an example to study the factors that influence the SSL model’s performance on lyrics recognition task.
\subsubsection{Effect of Language Model}
Building a good language model specifically for lyrics recognition is very challenging because lyrics often do not follow the standard grammar and have a lot of repetitions. We train an LSTM, transformer, and a 4-gram language model on the transcripts of the lyrics recognition training dataset and compare their performance in the decoding step.
As shown in Table \ref{table:3}, the LSTM and transformer achieve competitive WER, both significantly outperforming the 4-gram language model.
This is due to the fact that lyrics are too complex and arbitrary, but the 4-grams mainly learns local semantics. As a result, it fails to capture enough information to form a coherent model, which is why the LSTM and transformer language models perform much better as they can capture more of the global sequential relations. However, the lack of standard grammar in lyrics makes it difficult for the language model to abstract the high-level syntactic rules, hence the unique attention mechanism in the transformer architecture cannot be fully utilized. This is why transformer and LSTM language models have little difference in performance.

\subsubsection{Effect of Background Music on SSL Models}
We use the commercially recorded MusDB dataset to evaluate the effect of background music on SSL feature extraction. The MusDB dataset consists of high-quality recordings of 150 songs, each song has a version with background music and one without.
Because this dataset is small and to avoid the influence of downstream models on the results, we follow the idea of SUPER~\cite{yang2021superb} and only add a simple BiLSTM model as downstream of the SSL model. The encoder of the LSTM has 4 layers with 512 hidden units, and the decoder has 1 layer with 512 hidden units. 
For both of the audios with and without music, we train the model for 100 epochs and keep best five models. The HuBERT Large+LSTM model achieved a WER of 108.5 on the model with background music and a WER of 90.9 without background music.

\begin{figure}
 \includegraphics[width=.24\textwidth]{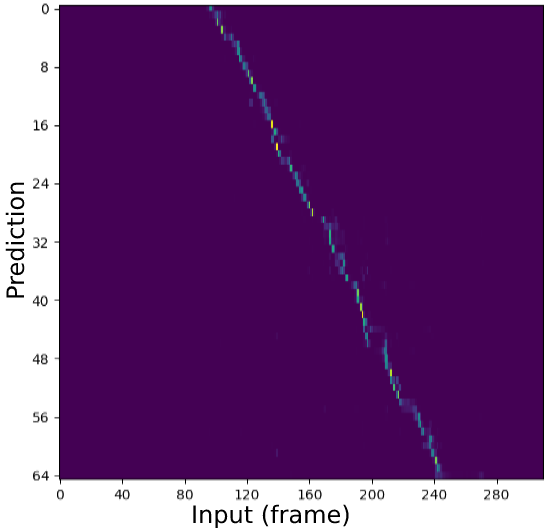}\hfill
\includegraphics[width=.24\textwidth]{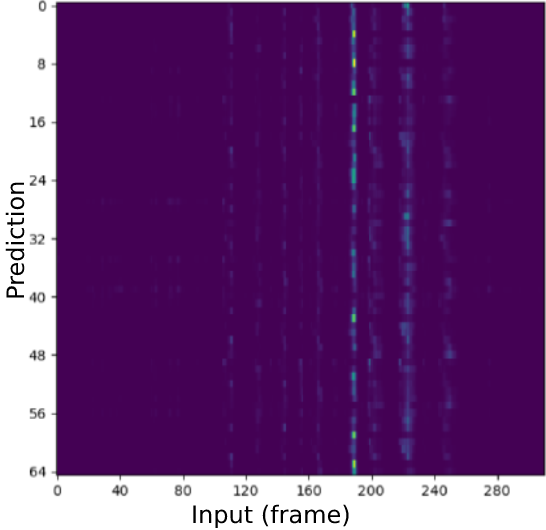}\hfill
\caption{Decoder attention weight graph: left:without Background Music right: with Background Music}\label{attention}
\end{figure}

Although training on background music characteristics helps acoustic models learn better lyrics alignment compared to background music suppression techniques~\cite{gupta2020automatic}, our results show that for SSL models, background music introduces noise that will impact SSL feature extraction from raw audio. 
This is because SSL models are trained by pure spoken utterances, but in songs, the sung voice and background music have correlated patterns. This correlation allows features extraction methods that extract information from the frequency domain using the envelop (MFCC, Mel Spectrum) to still be helpful, but SSL feature extractors operate on the time domain, in which case the overlap of the voice signal and music signal might result in completely different features.
Thus, the model tries to extract human acoustic features from the background music signals as well, resulting in poor performance. Furthermore, the presence of background music can overpower the human voice and weakens the strength of actual speech signals in the audio dataset, which results in all the outputs being predicted by a certain input frame as shown in Figure \ref{attention}.

\begin{table}
\begin{center}
\begin{tabular}{| l || c | c |}
 \hline
 HuBERT Large model WER Result & Test set & Dev set \\
 \hline\hline
 4-gram LM & 22.8& 23.0\\ 
LSTM LM & 18.9 & 18.6\\
Transformer LM & 17.2  & 18.1\\
\hline
\end{tabular}
\caption{HuBERT Large WER result for different language model on DAMP dataset}
\label{table:3}
\end{center}
\end{table}

%


\section{conclusion}
In this paper, we propose a lyrics recognition pipeline that extends upstream SSL feature extraction to downstream end-to-end ASR model for out-of-domain music signals. We established a new end-to-end model baseline for lyrics recognition on the Damp dataset, achieving a WER of 17.2 on the test set and 18.1 on the dev set. Even without lyrics language model trained on a large corpus, the performance improved by 5.23\% for the dev set and 2.4\% for the test set compared with the previous SOTA baseline systems. We demonstrate that the presence of background music complicates the feature extraction of SSL models trained on spoken utterances. We extend the research on SSL model out-of-domain adaptation and conclude that the loss of local information is the more probable cause of poor out-of-domain performance of SSL models. 

\clearpage

\begin{footnotesize}
\bibliography{reference}
\end{footnotesize}

\end{document}